\documentstyle[graphicx, amsmath, latin1]{mn}

\title{The impact of accretion disk winds on the X-ray spectrum of AGN: Part 1 - XSCORT}

\author[N.J.\,Schurch \& C.\,Done]
{N.J.\,Schurch$^{*}$ \& C.\,Done\\ 
Department of Physics, Durham University, South Road, Durham, DH1 3LE, UK \\
$^{*}$ nicholas.schurch@durham.ac.uk}

\date{}

\def\xmm{{\it XMM-Newton\/}}

\def\cha{{\it Chandra\/}}

\def\H0{{\rm ~km~s^{-1}~Mpc^{-1}}}

\def\etal{et al.~\/}

\def\la{\mathrel{\hbox{\rlap{\hbox{\lower4pt\hbox{$\sim$}}}{\raise2pt\hbox{$<$}}}}}
\def\ga{\mathrel{\hbox{\rlap{\hbox{\lower4pt\hbox{$\sim$}}}{\raise2pt\hbox{$>$}}}}}
\def\ls{\mathrel{\hbox{\rlap{\hbox{\lower4pt\hbox{$\sim$}}}\hbox{$<$}}}}
\def\gs{\mathrel{\hbox{\rlap{\hbox{\lower4pt\hbox{$\sim$}}}\hbox{$>$}}}}
\def\d25{D$_{\rm 25}$}

\def\.25{0.25 keV\thinspace}

\def\xst{{\small XSTAR}}

\def\xsc{{\small XSCORT}}

\begin{document}

\maketitle 

\begin{abstract}

The accretion disk in AGN is expected to produce strong outflows, in particular a UV-line driven wind. Several observed spectral features, including the soft X-ray excess, have been associated with the accretion disk wind. However, current spectral models of the X-ray spectrum of AGN observed through an accretion disk wind, known to provide a good fit to the observed X-ray data, are ad-hoc in their treatment of the outflow velocity and density of the wind material. In order to address these limitations we adopt a numerical computational method that links a series of radiative transfer calculations, incorporating the effect of a global velocity field in a self-consistent manner (\xsc). We present a series of example spectra from the \xsc ~code that allow us to examine the shape of AGN X-ray spectra seen through a smooth wind with terminal velocity of 0.3c, as appropriate for a UV-line driven wind. We calculate spectra for a a range of different acceleration laws, density distributions, total column densities and ionization parameters, but all these have sharp features that contrast strongly with both the previous 'smeared absorption' models, and with the observed smoothness of the soft X-ray excess.  This rules out absorption in a radiatively driven accretion disk wind as the origin of the soft X-ray excess, though a larger terminal velocity, possibly associated with material in a magnetically driven outflow/jet, may allow outflow models to recover a smooth excess. 

\end{abstract}


\section{Why are winds important?}
\label{1}

Mass outflows from accretion flows are observed to be ubiquitous features in accretion-powered astrophysical sources, with winds (and jets) being seen in systems as diverse as AGN and young stellar objects (e.g. the review by Livio 2003). These outflows are expected to carry both energy and angular momentum away from the associated accretion disk material and, as such, are intimately linked to the accretion flow ({\it e.g.} Blandford \& Payne 1982, Ampleford \etal 2007, Namouni 2007), however, the driving mechanism, particularly in the case of winds, may vary considerably between different sources. Different driving mechanisms give different physical properties (mass loss rate, outflow velocity, angular momentum) for the wind material, and thus the impact of the wind on the underlying accretion flow, as well as on the immediate environment, will be considerably different for different sources.

The most plausible physical mechanisms for launching a wind from an accretion disk around a black hole are thermal driving, radiation driving (both line \& continuum) and magnetic driving. Thermal driving results from irradiation of the outer disc by emission from the bright inner accretions flow and the coronal X-ray source. This creates a pressure difference between the irradiated surface of the disk and the vacuum above, which drives material up to larger scale heights. The Compton temperature is typically $\sim$10$^{7-8}$ K and is constant across the disk, however the escape velocity decreases with radius. Beyond a certain radius, then, the material driven out of the disk will have sufficient thermal energy to exceed the escape velocity of the system at that point. This launch radius, typically $\sim$10$^{4}$ Schwarzschild radii, is given by R$_{launch}$$>$10$^{10}$(M$_{BH}$/T$_{C,8}$) cm, where M$_{BH}$ is the mass of the black hole in solar masses and T$_{C,8}$ is the Compton temperature in units of 10$^{8}$K (Begelman \etal 1983).

Thermal winds should occur in {\it all} discs that have sufficient mass beyond the launch radius and so represent the {\it minimum} wind that is expected to be present in accreting sources. The mass loss rate of these winds can be dramatically enhanced if this material above the disc absorbs a substantial amount of radiation from the central source, with the imparted outward momentum giving rise to rapid radial acceleration of the wind material. Accounting for the radiation pressure, the effective gravity binding the material to the black hole is then reduced by a factor (1-f L/L$_{Edd}$). Here, L is the source luminosity, L$_{Edd}$ is the Eddington luminosity and f=($\sigma_{abs}$+$\sigma_{T}$)/$\sigma_{T}$ where $\sigma_{abs}$ and $\sigma_{T}$ are the true absorption and electron scattering cross-sections, respectively. For highly ionized material, f is dominated by the electron scattering cross-section which thus represents the {\it minimum} radiation pressure we expect to be exerted on any wind material, and gives rise to continuum radiatively driven winds around L/L$_{Edd}$$\sim$1. For less highly ionized material, f becomes dominated by the true absorption cross-section resulting in absorption line driving which can accelerate a wind at considerably lower $L/L_{Edd}$.

The potential of magnetic fields to drive a wind is difficult to quantify because the properties of such a wind will depend strongly on the field configuration (Blandford \& Payne 1982; Proga 2000; 2003). Some progress is being made towards understanding this ({\it e.g.} Hawley \& Krolik 2006, Kato 2007), but as yet there is no clear consensus, beyond the general agreement that winds will be substantially enhanced by magnetic fields, and that such a wind could be launched from any radius in the disk. This lack of diagnostic power means that magnetic driving can only be implied as the dominant wind driving mechanism for sources in which all the other mechanisms have been ruled out.

Moderate Eddington fraction ({\it i.e.} $\sim$0.1-0.5) stellar mass black holes and neutron stars, seen at high inclinations, often show high ionization blue-shifted absorption lines, typically associated with H-like and He-like iron. These lines imply the presence of columns of 10$^{23-24}$ cm$^{-2}$, outflowing at $\sim$200-500 km s$^{-1}$. Continuum radiation is ruled out as the dominant driving mechanism behind this material because the luminosities are considerably less than Eddington, while the high ionization state of the wind material rules out line driving. Instead the observed properties are generally well matched to thermal wind models ({\it e.g.} Woods \etal 1996), however there remains considerable debate in the community regarding the role of additional driving mechanisms, particularly concerning the role of magnetic fields (Miller \etal 2006, Netzer 2006).

The situation is slightly different in AGN due to the lower characteristic temperature associated with an accretion disk around a larger mass black hole. Coronal X-rays again push the material in the disk towards $\sim$10$^{7}$K, implying that a thermal wind could be launched from a similar gravitational radii in the disk as the thermal winds observed in X-ray binaries. However the lower disk temperature means that material lifted from the disk to larger scale heights, will not necessarily be as highly ionized as in the case of stellar mass black-holes, and that the accretion disk will produce a copious flux of UV photons. Such partially ionized material has huge opacity to UV and soft X-ray photons and resonance line absorption in such material can thus power a stronger wind, from smaller radii, than thermal driving. As such, UV line driving is expected to be the foremost physical mechanism for AGN accretion disk winds (Murray \& Chiang 1995, Proga \etal 2004). The predicted geometry for a UV-line driven wind has three distinct regions; a low density, completely ionized, funnel perpendicular to the disc, a slow, mostly neutral, outflow concentrated towards the plane of the disk, and a fast ($\sim$0.1-0.2c), partially ionized, outflow in-between, forming a bipolar funnel ({\it e.g.} Proga 2004).

Observationally, there is clear evidence for winds on large scales in AGN. In the X-ray band, a wide range of atomic transitions from partially ionized material (the warm absorber systems) are observed in many type-1 AGN ({\it e.g.} Kaastra \etal 2002, Blustin \etal 2005), however the inferred radii of this material places it in the region of the putative molecular torus, considerably further out than winds originating from an accretion disk, suggesting an origin for this material in a wind from an irradiated torus rather than from the accretion disc itself (Krolik \& Kriss 2001). Similarly, at higher inclination angles, intermediate type AGN show clear evidence for large column densities of moderately ionized absorbing material at large radii (most notably in NGC 4151 and Mrk 6 - Schurch \& Warwick 2002; Schurch, Griffiths \& Warwick 2006), however it is, again, unclear whether this material can be linked with winds from an accretion disk. In type-2 AGN, signatures from the accretion disc, and features from any associated wind, are overwhelmed by absorption in the neutral material of the obscuring torus.

There is also evidence for outflows with properties similar to those we expect from the putative UV-line driven accretion disk wind. BAL QSO's, for example, show highly ionized absorption lines in their optical and X-ray spectra, with relativistic outflow velocities of up to $\sim$0.2c (Weymann \etal 1991, Pounds \& Page 2006), and it has been suggested that the putative molecular torus may, in fact, be either the slowly outflowing, high column density, neutral material near the plane of the accretion disk, or an accretion disk wind that has stalled at large radii (Elvis 2000).

More indirect evidence for an outflow from the accretion disk comes from observations of the continuum spectral shape in high L/L$_{Edd}$ AGN, which typically shows a soft X-ray excess at energies below $\sim$1 keV ({\it e.g.} Porquet \etal 2004, Crummy \etal 2006). This smooth component shows no obvious atomic features, and can be modelled with a thermal continuum. Fitting a thermal continuum to a range of AGN X-ray spectra reveals that the soft excess has a characteristic `temperature' that is remarkably constant across a large number of sources, despite the large range of intrinsic source properties, making it unlikely that the soft X-ray excess is a true continuum component. The simplest mechanism for producing such a fixed 'temperature' component is atomic transitions in partially ionized material, in particular those associate with the strong jump in opacity between $\sim$0.7-3 keV caused by OVII/VIII and absorption in partially ionized iron. Fabian \etal (2002) suggested that the soft excess could be linked to partially ionized material in the optically thick accretion disc seen in reflection, with the relativistic motion of the disc smearing the characteristic sharp atomic features into a smooth, continuum-like, component. Alternatively, Gierli\'nski \& Done (2004 - hereafter GD04) suggested that the soft excess arises from absorption (and associated emission: Schurch \& Done 2006 - hereafter SD06) in partially ionized, optically thin material. In this case, the large velocity shear required to smooth the otherwise sharp atomic features implies that this material forms a wind and/or highly turbulent structure above the disc. 

Remarkably, this picture suggests that the accretion disk wind may be the dominant force shaping the X-ray spectrum we observe, and in some cases (notably 1H 0707-495; Sobolewska \& Done 2007), the impact of the wind may be so severe that it completely masks the underlying continuum shape. The model is successful in reproducing the soft X-ray excesses observed across a range of type-1 AGN (Middleton \etal 2007), together with their r.m.s. spectral variability (Gierli\'nski \& Done 2006). The GD04 and SD06 models are favoured by theoretical models of X-ray irradiated material in some sort of pressure balance, which show that the ionization instability limits the column density of material that can be present in the partially ionized state (Chevallier \etal 2006). This limited column is consistent with the optically thin absorption/emission models but is insufficient to dominate the entire photosphere of the accretion disc as required for the reflection origin (Done \& Nayakshin 2007). 

However, despite their success in modelling the data, the absorption/emission models have some serious deficiencies, particularly concerning the assumed velocity and density structure of the wind, which must introduce considerable scepticism in the validity of any conclusions based on these models (SD06).  

In this work we present a new approach to modelling the X-ray spectra of AGN observed through an accretion disk wind, that removes several of the limitations of previous models. Specifically, we develop a code (hereafter \xsc; \xst ~Simulation Chain for Outflows with Radiative Transfer) that allows us to impose an external velocity field on a photoionization code in a self-consistent manner, and show that this results in large changes in the shape of the resulting X-ray spectrum. In Section~\ref{2} we present the details of the \xsc ~code. In Section~\ref{3} we explore the behaviour of \xsc ~spectra based on the average wind parameters from the sample of PG quasars presented in Middleton \etal 2007. We examine how the model spectra change as the outflow velocity field, total column density and initial ionization parameter of the wind are varied. In Section~\ref{4} we examine the impact that a complex ionization structure in the wind has on the \xsc ~spectra and finally, in Section~\ref{5}, we present our conclusions.

\section{\xsc ~methodology}
\label{2}

The GD04 and SD06 models of AGN X-ray spectra observed through an outflow rely on radiative transfer codes to calculate the absorption and emission signatures associated with partially ionized material. The results of these calculations are then convolved with an external velocity field to produce a model X-ray spectrum that includes the impact of the accretion disk wind. The radiative transfer calculations are typically carried out under some simplifying assumptions, the two most serious of which are that the gas is (a) a continuous, thin, slab of gas and (b) has a uniform density throughout the material. In addition to these, the externally applied velocity field necessarily represents a vast simplification of what is likely to be a considerably complex physical situation. In particular, the GD04 \& SD06 models assume an external Gaussian velocity field, characterised by a very large dispersion ($\sigma\sim0.2-0.3c$). Such an external velocity field is only physically appropriate if the absorber consists of multiple, individual small clouds along the line-of-sight (l.o.s). In more continuous material, the velocity field sets the equivalent widths of the lines, so the photoionization cannot be calculated separately from the dynamics.

In order to address these limitations we adopt a numerical computation method that splits a single run of the radiative transfer code \xst ~into a series of multiple, linked, calculations that incorporates the effect of a global velocity field on the calculations in a self-consistent way. \xsc ~also allows us to relax several of the other assumptions present in the previous models, such as constant density, giving a considerably more realistic picture of the effects of an accretion disk wind on the X-ray spectra of AGN.

\subsection{Chaining \xst}
\label{2-1}

\xst ~calculates the absorption and emission spectrum that results from a stationary spherical shell of photoionized gas at a given radius from a given ionizing continuum, for a given a set of physical gas properties (density, abundance {\it etc}). \xst ~calculates the ionization balance and thermal equilibrium in a series of `sub-shells' of gas, propagating the input incident spectrum through these sub-shells in turn, calculating the ionization balance, temperature and opacities in each. Absorption and emission lines are calculated with an assumed Gaussian turbulent velocity broadening added in quadrature to the self-consistently calculated thermal line widths. The turbulent velocity is normally the dominant contributor to the line widths, controlling the total opacity in the absorption lines and hence the relation between the total column density of a given ion species and the observed line equivalent width ({\it i.e.} the curve of growth). It is important to note that the equivalent width of a line saturates as the line becomes optically thick, limiting its value, however a larger velocity width means that the intrinsic line width is larger and the line is thus less likely to become optically thick at its core. As a result, a large turbulent velocity, for a given column, leads to larger equivalent width absorption line. 

\xst ~currently only implements this turbulent velocity broadening on the absorption and emission lines, with the result that edges remain sharp despite the turbulent velocity being characteristic of the material as a whole (SD06). In fact, even if this were calculated self-consistently, the Gaussian nature of the turbulent velocity field would still be a poor representation of the velocity field in a relativistic outflow, where we would expect blue-shifted absorption along the l.o.s and both red and blue-shifted emission from all possible l.o.s. 

The computational method employed in \xsc ~is similar to that used internally in \xst ~in that we divide the total column up into a large number of shells, each with smaller column densities, and treat each shell in a separate, albeit linked, manner. In addition to this, however, the effects of the outflow velocity field are incorporated self-consistently for each shell in the chain. Specifically, we step through the shells in order using the relativistically velocity shifted output from the \xst ~calculation in the (i-1)$^{th}$-shell as the input for the \xst ~calculation in the i$^{th}$-shell, combined with a set of pre-defined gas parameters for the i$^{th}$-shell. Typically we specify five inputs for each shell in the \xsc ~chain; column density (N$_{H,i}$), gas density (n$_{i}$), ionization parameter ($\xi_{i}$), radial velocity ($v_{rad,i}$) and the input ionizing spectrum. 

\xst ~operates in the rest-frame of the absorbing material, so the incorporation of the outflow velocity requires that for each shell in the chain, the input luminosity, density and radius all be transformed into the rest-frame of the shell, using the appropriate relativistic relations, in order for \xst ~to calculate the ion populations correctly (see Appendix~\ref{app1-2} \& \ref{app1-3}). In addition, there is a subtlety in the way \xst ~treats line and continuum escape probabilities which means that it is not appropriate to simply take the output luminosity from a single \xst ~run and use this as the input luminosity to the next calculation. Instead we use the same photon field \xst ~propagates internally (Appendix~\ref{app1-1}) so that \xsc ~is consistent with a single, constant density, \xst ~run, in the absence of an external velocity field and constant gas density throughout the chain.

The velocity shear within each shell in the chain is characterised using the turbulent velocity parameter in \xst, $v_{turb,i}$. We set the turbulent velocity in shell $i$, v$_{turb,i}$ to be the standard deviation for a uniform distribution, with an additional component, $v_{t}$, representing the intrinsic real turbulent velocity of the material:

\begin{flushright}
\begin{minipage}{80mm}
\begin{eqnarray}
v_{turb,i} & = & (v_{rad,i}-v_{rad,i-1})/\sqrt{12}\,+\,v_{t}
\label{eq1}
\end{eqnarray}
\end{minipage}
\end{flushright}

For the spectra presented in Sections~\ref{3} \& \ref{4} the intrinsic turbulent velocity, $v_{t}$, is set at 100 km s$^{-1}$. The derivation of $v_{turb,i}$ is given in detail in Appendix~\ref{app1-4}, but intuitively, it is clear that the Gaussian dispersion must be less than half the velocity shear because a Gaussian distribution is symmetric about the mean. 

\subsection{Calculating the transmitted and emitted spectra}
\label{2-2}

The transmitted spectrum from shell $i$ is calculated by transforming the transmitted spectrum from the $i-1^{th}$ shell, in the observers frame, into the rest-frame of shell $i$, absorbing the spectrum with the rest-frame opacities calculated by \xst, and then transforming the newly absorbed spectrum back into the observers frame. Iterated over all the shells in the chain this then produces the total transmitted spectrum from the outflowing wind. 

Once in the rest-frame of shell $i$, the transmitted spectrum is calculated in a similar fashion to the calculation \xst ~performs internally (see Appendix~\ref{app1}). At this point, however, we deviate from our requirement of consistency with a single \xst ~run by including a simple treatment of electron scattering (something \xst ~does not currently incorporate), scattering flux out of the transmitted (l.o.s) spectrum and into the emitted (global) spectrum. We follow the notation used in the \xst ~manual$^{*}$\footnotetext{\hspace{-2mm}$^{*}$\hspace{1mm}http://heasarc.gsfc.nasa.gov/docs/software/xstar/docs/ html/xstarmanual.html} and denote all specific luminosities (units of erg/s/erg) as $L_{\epsilon}$. The transmitted spectrum from shell $i$, denoted as $L^{5}_{\epsilon,i}$ in the \xst ~manual, is then given by:

\begin{flushright}
\begin{minipage}{80mm}
\begin{eqnarray}
L^{5}_{\epsilon,i} & = & L^{5}_{\epsilon,i-1}\,e^{-(\tau_{tot,i}+N_{H,i}\sigma_{T}\,)}
\label{eq2}
\end{eqnarray}
\end{minipage}
\end{flushright}

\hspace{-6mm}where $L^{5}_{\epsilon,i-1}$ is the output transmitted spectrum from the previous shell, transformed into the rest-frame of shell $i$ and $\tau_{tot,i}$ is the sum of the bound-bound, bound-free and free-free optical depths. $\tau_{T}=N_{H,i}\sigma_{T}$ is the optical depth to electron scattering, where $N_{H,i}$ is the column density of the shell and $\sigma_{T}$ is the Thompson cross-section. We note that this is simplistic treatment of electron scattering, neglecting the effects of absorption, multiple scatterings, down-scattering and the Klein-Nishina cut-off. The latter two of these effects are largely unimportant at energies $<$100 keV, while the contribution of multiple scatterings depends on N$_{H,i}$ but has less than a 10\% effect on the spectral shape for column densities of $\sim$10$^{23}$ cm$^{-2}$.

While the impact of the velocity on the transmitted spectrum is relatively simple to calculate, the impact on the emission spectrum is somewhat less straightforward due to the global nature of the wind material. We assume that the emission originates from a spherically symmetric wind, and thus the total emission spectrum will include a contribution from every possible l.o.s. We divide the spherical shell, $i$, into segments of solid angle $d\Omega=\sin\theta\,d\theta d\phi$. The radial emission spectrum from each solid angle segment is then the emission spectrum output from \xst, re-normalised by $d\Omega/4\pi$. Finally, the total l.o.s emission spectrum from shell $i$ is then the sum of the emission spectra from each solid angle segment, shifted and beamed by the component of the outflow velocity of the segment along the l.o.s. The emission from each solid angle segment will be absorbed by wind material along its l.o.s to the observer (which will vary from almost nothing for material in the outer shells on the near side of the source, to much more than the continuum l.o.s absorption for material on the far side of the source) however, calculating the absorption for the emission spectrum increases the computational complexity of the code enormously, and dramatically increases the run-time of the program, so for the work presented here we restrict ourselves to calculating the unabsorbed total emitted spectrum and stress that this is an overestimate of the emission from the wind. 

The emission spectrum is also sensitive to both the local and global covering fraction of the wind material. Reducing the local covering fraction used by \xst ~({\it i.e.} whether the material in the immediate environment of each l.o.s is clumpy or uniformly smooth) can reduce the overall normalisation of the emission spectrum by a factor $\sim$2, but does not systematically change the shape of the emitted spectrum (we hardwire the local covering fraction $C_{f,local}$=0.001). Alternatively, imposing a large scale geometry on the wind material (evacuating a bi-conical structure along the spin axis of the central black hole, for example) can potentially have a considerable effect, but only on the emission spectrum. In the case of a bi-conical structure, a continuous region of solid angle is removed as a source of emission, resulting in a change in the detailed shape of individual emission features as well as a reduced overall normalisation. In the most extreme case, where the only remaining material is distributed in a thin equatorial disk, this results in the classic double-peaked emission lines familiar from accretion disk studies ({\it e.g.} Chen \& Halpern 1989). The model spectra presented are calculated for a global covering fraction, $C_{f,global}$=1, and thus represent the maximum possible contribution of the emission component to the total spectrum.

The rest-frame emitted spectrum is calculated in a similar fashion to the calculation \xst ~performs internally (see Appendix~\ref{app1}), however we again deviate from our requirement of consistency with a single \xst ~run, by adding the electron scattered component to the genuine emission spectrum. Again, using the same notation as the \xst ~manual, the total rest-frame emission spectrum from shell $i$, denoted $L^{7}_{\epsilon,i}$, is thus given by:

\begin{flushright}
\begin{minipage}{80mm}
\begin{eqnarray}
L^{7}_{\epsilon,i}{\hspace{2mm}}= & L_{\epsilon,bb,i}\,+\,L_{\epsilon,bf,i}\,+\,L_{\epsilon,ff,i} & \nonumber\\
 & +\,L^{5}_{\epsilon,i-1}\,e^{-(N_{H,i}\sigma_{T}\,)} & \mid\,_{outward}
\label{eq3}
\end{eqnarray}
\end{minipage}
\end{flushright}

\hspace{-6mm}where $L_{\epsilon,bb,i}$, $L_{\epsilon,bf,i}$ \& $L_{\epsilon,ff,i}$ are specific luminosity spectra from bound-bound, bound-free and free-free atomic transitions. We note again that $\tau_{T}=N_{H,i}\sigma_{T}$ is the optical depth to electron scattering, where $N_{H,i}$ is the column density of shell $i$, $\sigma_{T}$ is the Thompson cross-section and thus the final term in this equation represents the flux scattered into our l.o.s through electron scattering.

The transmitted and emitted spectra are output on the same energy grid as that used by \xst, namely a logarithmic energy grid spanning E=0.1-10$^{6}$.

\subsection{Ionization Structure}
\label{2-3}

\begin{figure}
\centering
\begin{minipage}{85 mm} 
\centering
\includegraphics[height=8 cm, angle=270]{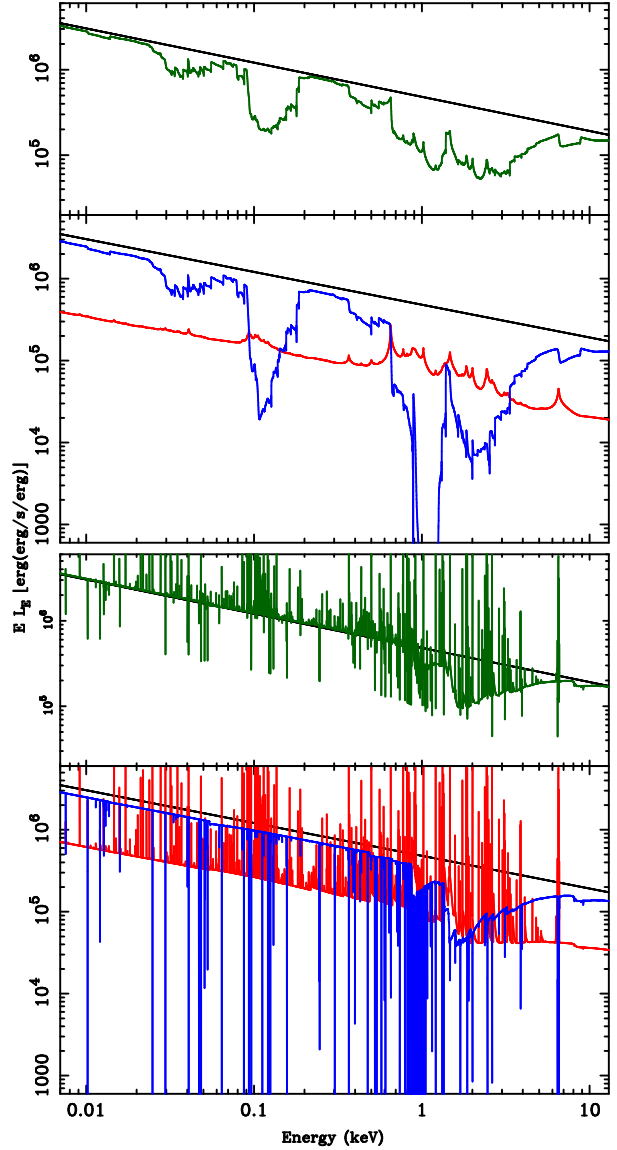}
\caption{{\small {\it Top Panel:} The UV/X-ray spectra ($E\,L_{E}$) that results from an \xsc ~run with a linearly accelerating velocity field. {\it Bottom Panel:} The X-ray spectra that results from an \xsc ~run without an external velocity field. {\it Upper:} The total spectrum (dark green) and the input power-law continuum (black). {\it Lower:} The transmitted spectrum (blue), the emitted spectrum (red) and the input power-law continuum (black).}}
\label{xscfull}
\end{minipage}
\end{figure}

\begin{figure}
\centering
\begin{minipage}{85 mm} 
\centering
\vbox{
\includegraphics[width=8 cm, angle=270]{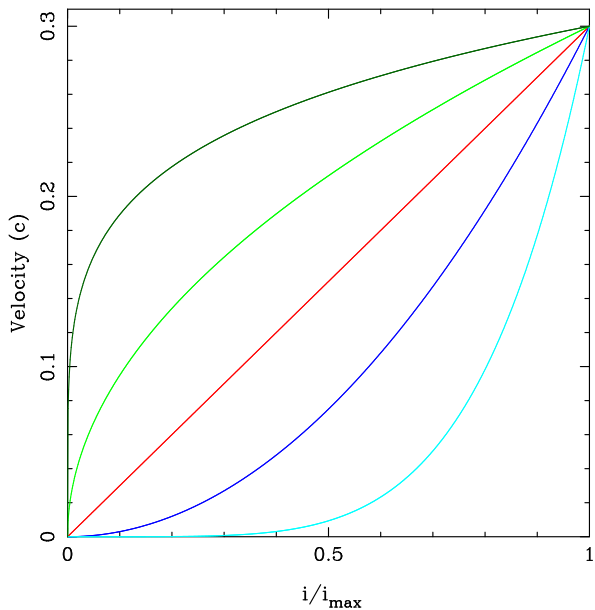}
}
\caption{{\small Velocity as a function of step through the \xsc ~chain. The red line corresponds to linear acceleration ($p$=1). The dark and light green lines have $p$=0.2 \& 0.5, respectively and represent the situation where the majority of the acceleration of the wind happens near the launch radius. Conversely, dark and light blue lines have $p$=2 \& 5 and represent the situation where the majority of the acceleration of the wind happens far from the launch radius. in each case, the launch velocity is 0 and the outer-edge velocity is 0.3c.}}
\label{vlaw}
\end{minipage}
\end{figure}

The relativistic nature of the outflow has a strong impact on the ionization structure of the wind. For an outwardly accelerating wind, the input spectrum for shell $i$ ($L^{1}_{\epsilon,i-1}$ - see Appendix~\ref{app1-1}) will be both red-shifted and beamed (away) relative to shell $i-1$. These velocity transformations (Appendix~\ref{app1-3}) have a strong impact on the luminosity of the input ionizing spectrum (a factor of $\sim$3 for terminal velocities of $\sim$0.3c), reducing the ionization state of the wind material at each step. The density of the wind material along the l.o.s is also affected by the wind outflow velocity. Specifically, the density defined in the observers frame is related to the density in the rest frame of the shell through a length contraction based on the l.o.s outflow velocity. The density is reduced by a factor of 1/$\gamma$, where $\gamma=(1-v_{i}^{2}/c^{2})^{1/2}$ and v$_{i}$ is the l.o.s component of $v_{rad,i}$, the radial velocity of shell $i$. The impact of the length contraction is relatively small ($\sim$5\% for v$_{out}=0.3c$) in contrast to the impact that the transformed luminosity has on the ion populations. 

Additionally, imposing an external density structure on the wind will impact on the ionization structure of the wind. In order to understand the properties of the \xsc ~models, in Section~\ref{3} we impose a density structure on the wind material such that the changes in density counteract the red-shift and beaming effects that affect the input ionizing spectrum. To first order this results in a constant ionization state throughout the material; a situation that is broadly familiar from static models of ionized material. The remaining change in the ionization parameter (typically $\la$3\%) as we move through the wind material reflects only the decrease in luminosity due to absorption by the wind material and the increase in distance of each shell from the ionizing source, due to the geometric width of the wind.

The ionization instability provides a natural physical mechanism for producing a wind with a constant ionization state structure. Moving through a slab of highly ionized gas, the transition from highly ionized material to low ionization material occurs over a small range of radii. As the ionization state of the material drops, however, the material will cool and collapses, increasing its density which, in turn, drives the ionization state down even further. This runaway process results in almost neutral, dense, clumps of material, however their volume filling fraction is small and as a result these clumps are not expected to have a large impact on the X-ray spectrum. The dominant contribution to the X-ray spectra shape is, then, from the relatively thin layer of partially ionized material with (almost) constant ionization state (Chevallier \etal 2006). 

This scenario does not, however, account for the impact of the outflow velocity. The clumps will collapse on timescales of approximately sound crossing time of the clump. If the wind outflow velocity is considerably greater than the sound speed in the gas before it begins to collapse, then these clouds may not be able to collapse effectively. In Section~\ref{4} we relax the previous constraint on the density structure of the wind and explore winds with a density structure that thus have wide range of ionization parameters through the wind material. We take the simplest possible approach to the density structure of the wind and treat only structure in the radial direction. Because the transmitted spectrum only samples a pencil-beam l.o.s, it is only sensitive to the radial density structure of the wind and is insensitive to the covering fraction or the filling fraction of the gas. In contrast, the emitted spectrum will be strongly affected by the 3D density structure of the flow, however the emission is not the dominant component of the total spectrum except in calculations with particularly high column densities, or low ionization states. Calculating the impact of a detailed 3D structure on the emitted spectrum is beyond the scope of the work presented here and, for the majority of the spectra presented here, is not expected to strongly impact the shape of the resulting spectra. We stress again that the spectra presented here have the maximum possible contribution of the emission component to the total spectrum (see Section~\ref{2-2}) and as such should be treated as an upper limit in this regard.

\section{Behaviour of the output spectra from \xsc}
\label{3}

\begin{figure*}
\centering
\begin{minipage}{170 mm} 
\centering
\includegraphics[width=119 mm, angle=270]{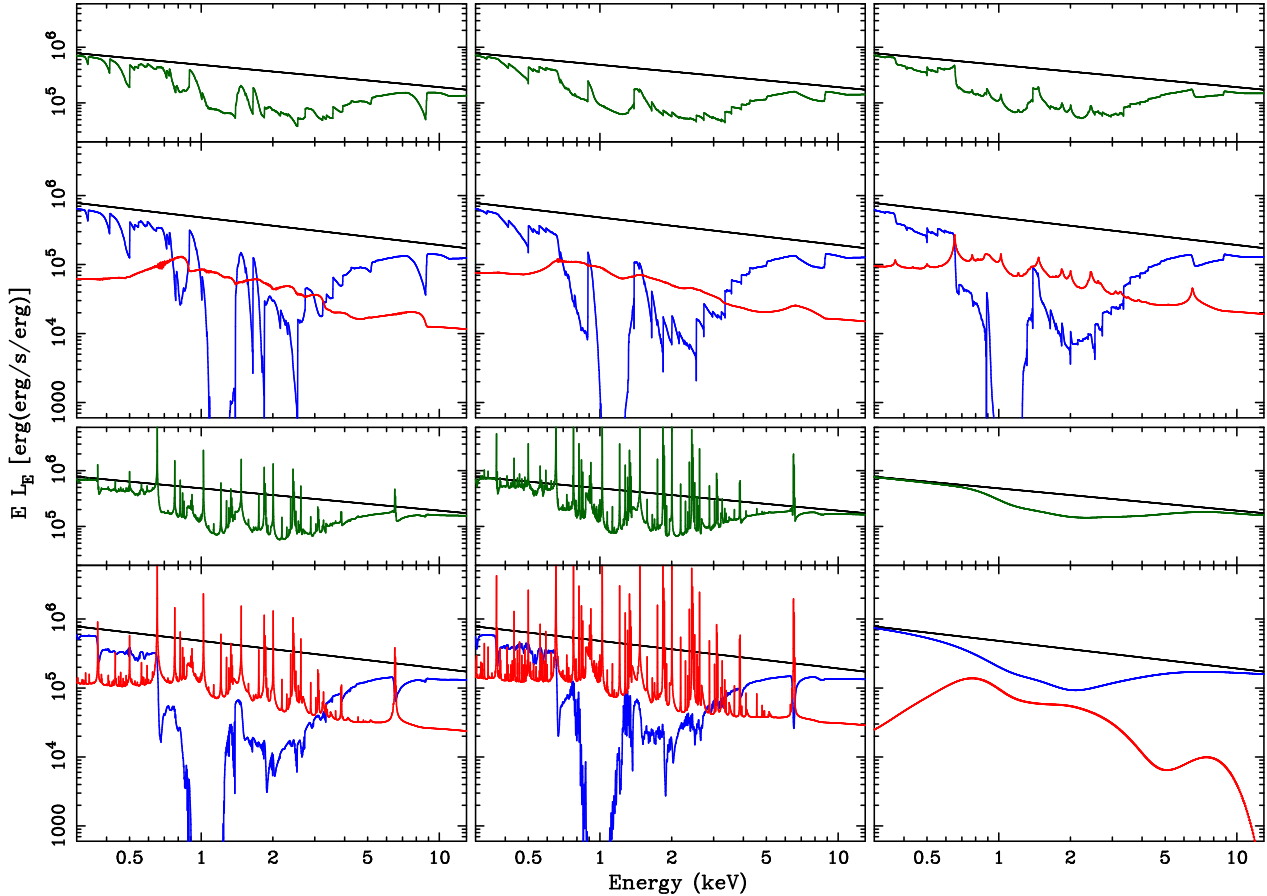}
\caption{{\small The 0.3-13 keV X-ray spectra ($E\,L_{E}$) that result from \xsc ~runs with different $p$ velocity laws: Panels 1-3 (Top, left to right): $p$ = 0.2, 0.5 \& 1.0 (linear acceleration). Panels 4-6 (Bottom, left to right): $p$ = 2.0, 5.0 \& the {\it Swindabs+Swindemm} model from Schurch \& Done (2006). All other parameters are fixed at the standard values. {\it Upper:} The total spectrum (dark green) and the input power law continuum (black). {\it Lower:} The transmitted spectrum (blue), the emitted spectrum (red) and the input power-law continuum (black). We note that Middleton \etal (2007) fit only smeared absorption to their sample, ignoring the associated emission component. Thus, the blue line in the bottom right panel represents the smeared absorption Middleton \etal fit. We stress that absorption of the emission spectrum in subsequent shells is neglected throughout this work and that all the spectra are calculated for a global covering fraction of $C_{f,global}$=1. Thus, the emission spectra presented here represent the maximum possible contribution from emission to the total spectrum.}}
\label{vlawspect}
\end{minipage}
\end{figure*}

In this section we examine the behaviour of the \xsc ~models as key parameters are varied, to highlight the impact that specific parameters have on the overall shape of the X-ray spectrum seen through an accelerating wind. We use a power law continuum as the initial ionizing spectrum and we define a standard set of parameters based, in part, on the set of `average' AGN wind properties from Middleton \etal 2007.

Middleton \etal (2007) analyse X-ray observations of a moderate sized sample of nearby, bright, PG quasars and NLS1s, and show that the X-ray spectra of this sample can be equally well fit by partially ionized, optically thin, smeared, material seen in absorption, or by optically thick, partially ionized, material seen in reflection (Crummy \etal 2006). The average absorption in the Middleton \etal sample has an ionization of log$_{10}$($\xi$)=3, a column density of N$_{H}$=3$\times$10$^{23}$ cm$^{-2}$ and an outflow velocity of $\ga$0.3c, although there is a considerable range across the sample in each of these parameters (factors of $\ga$10 in the first two cases and a factor of $\sim$ three in the latter cases).   The standard set of parameters is then: a 100 step chain, Log$_{10}$($\xi$)=3, N$_{H,tot}$=3$\times$10$^{23}$ cm$^{-2}$, an initial gas density of 10$^{12}$ cm$^{-3}$ and linear acceleration between 0 and 0.3c. In each of the following subsections we then vary one of these parameters.  With these parameters, the radial depth of the winds simulated with these parameters is small compared to the launch radius. However the physics that governs the shape of the output X-ray spectra depends only the distribution of column density and atomic species as a function of outflow velocity. Under the assumption that the gas is smooth, reducing the gas density in order to simulate a geometrically thicker wind, would thus force the launch radius of the wind out to large radii in order to have the given ionization parameter but the resulting spectrum would stay approximately the same.   These parameters are also similar to those resulting from  hydrodynamic simulations of accretion disc winds in AGN, though these also have a complex, filamentary structure which is beyond the scope of this paper. Nonetheless, these filaments have gas densities $\sim$10$^{9-13}$ cm$^{-3}$ and maximum velocities of $\sim 0.2c$, with the majority of the acceleration occurring close to the base of these structures (Proga, Stone \& Kallman 2000; Proga \& Kallman 2004).

\subsection{The outflow velocity field}
\label{3-1}

We examine the effect of a basic accelerating velocity law on the resulting X-ray spectrum, adopting a simple $p$-type form; {\it i.e.} the radial velocity of shell $i$ is given by:

\begin{flushright}
\begin{minipage}{80mm}
\begin{eqnarray}
v_{rad,i} & = & v_{l} + (v_{o}-v_{l})\left(\frac{i}{i_{max}}\right)^{p}
\label{eq4}
\end{eqnarray}
\end{minipage}
\end{flushright}

\hspace{-6mm}where $v_{l}$ and $v_{o}$ are the launch and terminal velocities, respectively, and $i_{max}$ is the number of steps in the \xsc ~chain. We note that for $p$=1, equation~\ref{eq4} produces a linear acceleration through the wind material. 

Figure~\ref{xscfull} shows an example of the X-ray spectrum that results from an \xsc ~run simulating an accelerating wind, over the 7~eV to 13~keV range, and a spectrum that does not include an external velocity field. The contrast between the two spectra highlights the massive impact that the velocity field has on both the transmitted and emitted spectrum. The \xsc ~spectrum is clearly dominated by two prominent absorption regions around $\sim$0.06 and $\sim$1.5 keV. The first of these absorption conglomerates is the result of low ionization species (C{\small III}, C{\small IV}, N{\small III}, Fe{\small V} {\it etc}) and falls well outside the band-pass of current X-ray observatories. Interestingly, the UV absorption is similar to the broad absorption lines that are characteristic of BAL QSO's, a possibility that we will examine in future work by combining \xsc ~with hydrodynamic simulations of accretion disk winds. The second of the absorption troughs is the result of higher ionization species (N{\small VI}, O{\small VII}, O{\small VIII}, Fe{\small XVII}, {\it etc}) and occurs in a region that is well covered with the current generation of X-ray observatories. While the stationary model has the continuum absorption dominated by bound-free edges, with individual absorption lines superimposed on this, the increased velocity shear in the wind model results in the resonance lines completely dominating the opacity. This produces a dramatic increase in the apparent absorption of the continuum spectrum when a velocity field is included in the calculations. For the remainder of this work we concentrate of the 0.3-13~keV band-pass observable with \xmm ~\& \cha, although we note that the $\sim$0.02-0.09~keV absorption feature may well be detectable in current, or future, far-UV spectra of AGN.

Figure~\ref{vlaw} shows the wind velocity as a function of fraction through the total processing chain, for several values of $p$. There is a considerable range in the mean velocity of the simulated winds in Figure~\ref{vlaw} ($<v>$=$v_{o}/(p+1)$=0.05 \& 0.25c for $p$=5 \& 0.2) however the velocity dispersion is similar in each case ($\sigma$=$<v>p/(2p+1)^{1/2}$=0.07c \& 0.04c for both $p$=5 \& 0.2). Figure~\ref{vlawspect} (panels 1-5) shows the model X-ray spectra from \xsc ~with these velocity laws, highlighting the range of spectral shapes that result. To zero-th order the transmitted component of the X-ray spectrum remains largely insensitive to the details of the velocity field because this component is dominated by the effects of velocity dispersion. In contrast, the emission spectrum, which includes contributions from red-shifted material from the far side of the wind as well as blue-shifted material from the near side of the wind, is sensitive to a combination of both the mean velocity and velocity dispersion. 

\begin{figure}
\centering
\begin{minipage}{85 mm} 
\centering
\vbox{
\includegraphics[width=12 cm, angle=270]{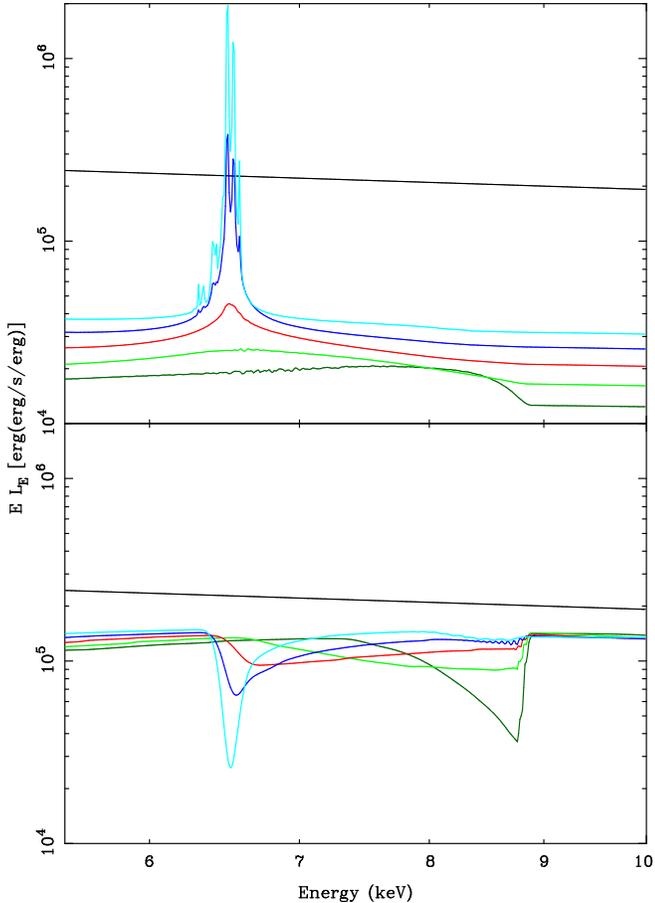}
}
\caption{{\small Structure in the iron K$\alpha$ region as a function of velocity law. The dark green, green, red, dark blue \& light blue lines correspond to the velocity laws shown in Figure~\ref{vlaw} ($p$=0.2, 0.5, 1.0, 2.0 \& 5.0, respectively). The black line is the input power-law continuum. {\it Upper Panel}: Emission spectra. {\it Lower Panel}: Transmission spectra. High $p$-values result in narrow iron line and iron edge features at almost rest-frame energies. Low $p$-values, however, result in considerable smearing, particularly of the iron line, producing features with maximum emission and absorption at energies that correspond to the maximum velocity of the outflowing material.}}
\label{vironedge}
\end{minipage}
\end{figure}

Detailed differences in the shape of isolated emission, and absorption, features such as the iron K$\alpha$ line and edge (Figure~\ref{vironedge}) explicitly reveal the distribution of the column density of a specific ion with velocity. At high $p$-values ({\it e.g.} $p$=5) a considerable amount of the wind material is present in regions where the velocities and, crucially, the velocity gradient are both small. For absorption features, sensitive only to the velocity dispersion, most of the opacity at high $p$-values occurs at near-rest-frame energies with an tail to higher energies. The emission features from a wind with a high $p$-value are not highly smeared and are imprinted on the spectrum as narrow features with maximum absorption and emission at almost rest-frame energies. Conversely, at low $p$-values ({\it e.g.} $p$=0.2) a considerable amount of the wind material is present in regions where the velocities and, crucially, the velocity gradient are both large. For absorption features, this results in the majority of the opacity occurring at energies corresponding to the mean blue-shift of the material, with a tail to lower energies. For the emission features this results in strongly broadened features with maximum emission at energies that, again, correspond to the mean blue-shift of the material (Figure~\ref{vironedge}, upper panel).

This highlights an issue concerning the strong iron K$\alpha$ features observed in some nearby Quasars. These model demonstrate that absorption in an accretion disk wind {\it can} make the sharp drop seen around 7 keV in some AGN, as long as the majority of H-like iron column is at low velocity (Done \etal 2006). However, the emission from this material is then far too narrow to create a smooth P-Cygni profile such as we might expect from a wind. The impact of the emission can be reduced by reducing the covering factor of the wind, in which case the profile is then dominated by the absorption, but we caution that our emission model does not include resonant line scattering which will enhance the line emission in the emitted spectrum.  A full treatment of scattering, absorption and emission of resonance lines is extremely complex and beyond the scope of this work (Matt \etal 1996).

\begin{figure}
\centering
\begin{minipage}{85 mm} 
\centering
\vbox{
\includegraphics[height=8 cm, angle=270]{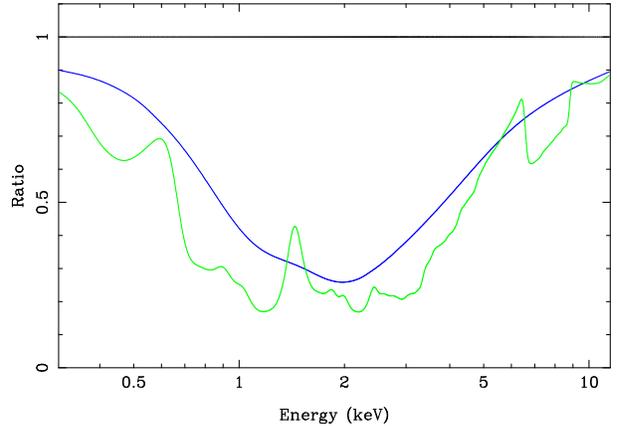}
}
\caption{{\small The ratio of the standard linearly accelerating \xsc ~model, convolved with the \xmm ~EPIC PN instrument response, to a simple power-law, also convolved with the instrument response (green). Also shown is the ratio of the smeared ionized absorption used in Middleton \etal (2007) convolved with the same instrument response, to a simple power-law (blue).}}
\label{instrresp}
\end{minipage}
\end{figure}

More importantly, these models show a key difference from both the observed shape of the soft X-ray excess and the simplistic, Gaussian-smeared wind models. Firstly, for the same column density of material, all the accelerating wind spectra show considerably more absorption than the spectra from a slab ionization model convolved with a Gaussian velocity field. The proper treatment of velocity in these models means that the absorption lines stay on the optically thin part of the curve of growth and their equivalent width, for the same column, is thus considerably greater. We note that it is not quite equivalent to simply use a lower column in the \xsc ~models, because the impact of the wind velocity enhances the line absorption over the bound-free edges. 

\begin{figure*}
\centering
\begin{minipage}{170 mm} 
\centering
\includegraphics[width=119 mm, angle=270]{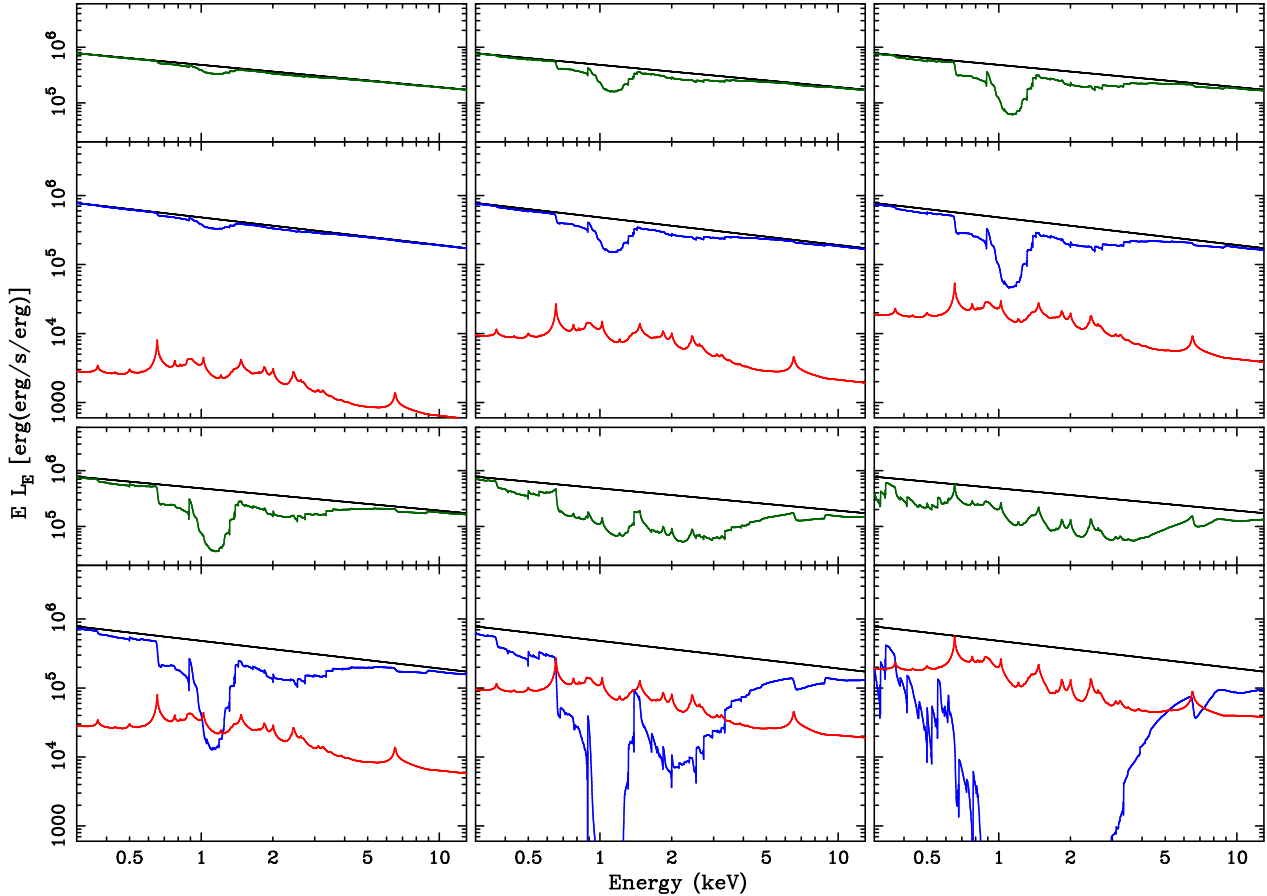}
\caption{{\small As for Figure~\ref{vlawspect} but for \xsc ~runs with different total column densities: Panels 1-3 (Top, left to right): $N_{H}$ = 9$\times$10$^{21}$, 3$\times$10$^{22}$ \& 6$\times$10$^{22}$ cm$^{-2}$. Panels 4-6 (Bottom, left to right): N$_{H}$ = 9$\times$10$^{22}$, 3$\times$10$^{23}$, 6$\times$10$^{23}$ cm$^{-2}$. All other parameters are fixed at the standard values.}}
\label{NHspect}
\end{minipage}
\end{figure*}

Even more importantly, the accelerating wind spectra have sharp features in the total spectrum that are not present in the GD04 \& SD06 models. The presence of narrow features is a product of the reasonably small velocity dispersion in the velocity laws shown, compared to the $\sim$0.3c velocity dispersion required by the GD04 and SD06 models. The sharp features predicted by the \xsc ~models are observable even with current high resolution X-ray CCD detectors. Figure~\ref{instrresp} shows the total spectrum from the standard linearly accelerating \xsc ~model convolved with the \xmm ~EPIC PN instrument response divided by a simple power-law model, also convolved with the instrument response. Also shown is the ratio for the smeared ionized absorption used in Middleton \etal (2007).  The overall shape of the smeared ionized absorber model and the \xsc ~wind models are broadly similar, albeit that the \xsc ~wind model shows greater absorption, for the same column density. However, the sharp features at 0.6 \& 1.4 keV are both clearly visible in the \xsc ~model, as is the strong drop from H-like iron K$\alpha$ absorption lines around 7 keV. The strong, line-like, feature at $\sim$1.4 keV is, in fact, not a line (although there is a considerable contribution to the flux at these energies from line emission). It is primarily the result of considerable broadened absorption at energies directly above and below 1.4 keV, associated with ionized oxygen and iron respectively. While it is currently not possible to fit the \xsc ~models to observed X-ray spectra, the presence of this dramatic feature suggests that the \xsc ~models presented here will be unable to reproduce the observed smooth featureless soft X-ray excess. Again, we stress that the effect of absorption on the emission spectrum has been neglected in the calculations presented here, and that this will imprint additional features on the emission spectrum and thus have a considerable impact on the total spectrum. However we note that these features will be similar in character to the features already evident in the transmitted spectrum and, as such, will not remove the sharp features that are present in the current total spectrum. In fact, in regions of the spectrum where the (comparatively) smooth emitted spectrum dominates the total spectrum, incorporating the absorption on the emission spectrum will introduce additional sharp features into the resulting total spectrum! Clearly this will not result the \xsc ~models being more compatible with the simplistic smeared ionized absorption models or the observed X-ray spectra of AGN.

\begin{figure*}
\centering
\begin{minipage}{170 mm} 
\centering \includegraphics[width=119 mm, angle=270]{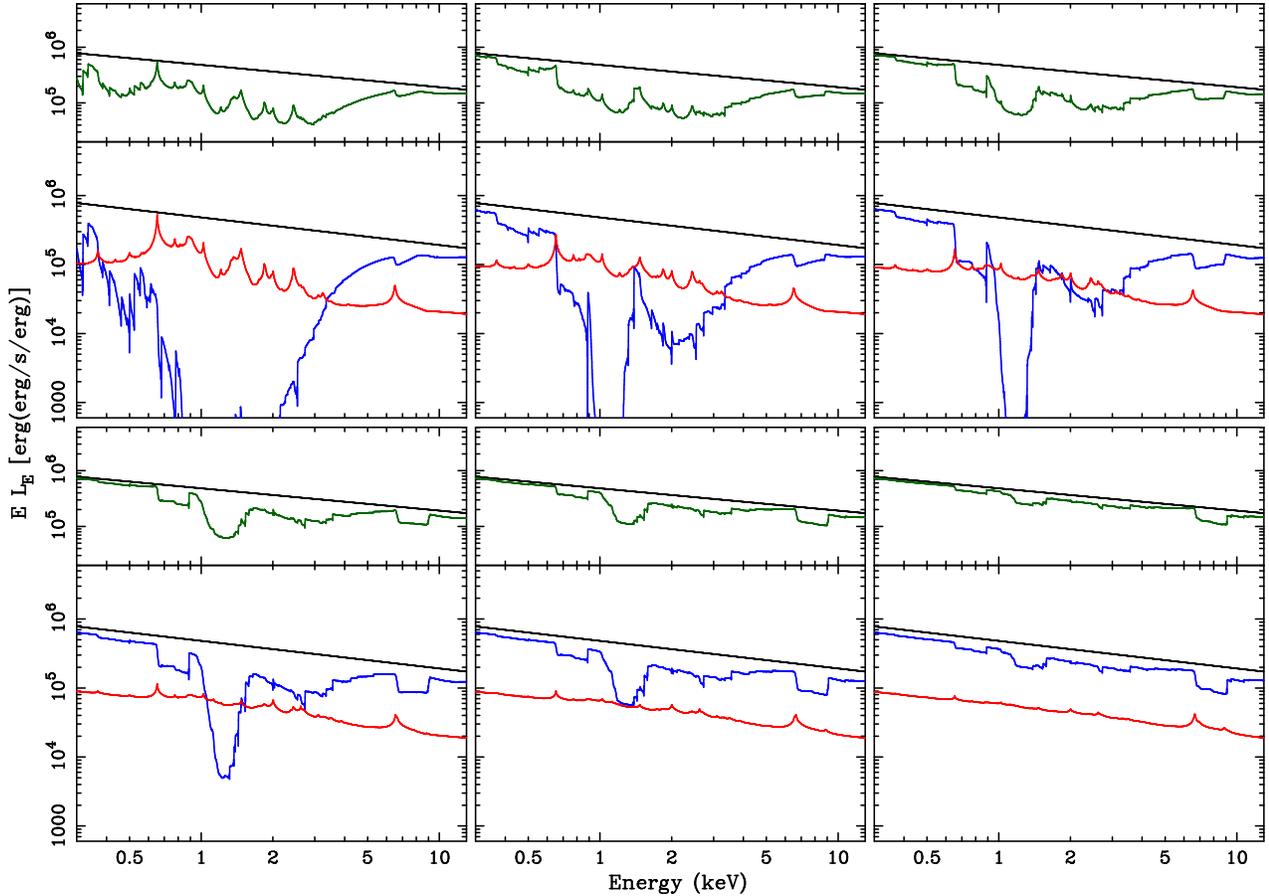}
\caption{{\small As for Figure~\ref{vlawspect} but for \xsc ~runs with different initial ionization states: Panels 1-3 (Top, left to right): log$_{10}$($\xi$) = 2.75, 3.0 \& 3.25. Panels 4-6 (Bottom, left to right): log$_{10}$($\xi$) = 3.5, 3.75 \& 4.0. All other parameters are fixed at the standard values.}}
\label{chispect}
\end{minipage}
\end{figure*}

The sharp features, particularly the feature at $\sim$1.4 keV, can be removed, however, by increasing the terminal velocity of the wind, increasing the energy range over which the prominent absorption troughs have a strong impact. The mean velocity dispersion required, $\sim$0.3c, corresponds to a terminal velocity of $\sim$c, implying an association with material in the jet rather than material in an accretion disk wind.

\subsection{Total column density}
\label{3-2}

Figure~\ref{NHspect} shows the X-ray spectra from winds with a range of total column densities.

At low column densities ($<$6$\times$10$^{22}$ cm$^{-2}$) the total spectrum is dominated by the transmitted spectral component. Above these columns the absorption is significant enough that the emission spectrum begins to dominate regions of the spectrum. At very high column densities ($>$few$\times$10$^{23}$ cm$^{-2}$) the absorption is so significant that the $<$10 keV spectrum becomes completely dominated by the emission component. We stress, again, that the emission spectrum calculated by \xsc ~is, currently, not self-consistently absorbed by the wind material, resulting in an emission spectrum that is somewhat brighter than we would otherwise expect, particularly at high column densities (see Section~\ref{2-2}). For most of the spectra shown here, the ionization parameter does not change appreciably across the thin wind currently simulated by the \xsc ~chain. As such, the ionic species present in the wind do not change considerably as a function of radius and, thus, the absorption and emission features do not change character significantly, despite increasing considerably in strength. For thicker winds (or columns $>$few$\times$10$^{23}$ cm$^{-2}$) the ionization parameter will change considerably across the width of the wind and the resulting changes in the ionic species with radius results in considerable changes in the shape of both the emitted and transmitted spectrum (see Section~\ref{4}).

\subsection{Ionization parameter}
\label{3-3}

Figure~\ref{chispect} shows examples of the X-ray spectra from winds with a range of initial ionization states.

At low ionization states (Log$_{10}$($\xi$)$\leq$2.5) the X-ray spectrum is dominated at high energies by the transmitted spectrum and at low energies by the emission spectrum (again, somewhat overestimated by lack of self absorption). Above (Log$_{10}$($\xi$)$\geq$2.5) the total spectrum becomes increasingly dominated by the transmitted component as there is less absorption and hence less emission, and the features imprinted on the X-ray spectrum become dominated by high ionization species. In particular, the iron K$\alpha$ absorption line becomes stronger as more of the column is ionized to He- and H-like iron.

\begin{figure}
\centering
\begin{minipage}{85 mm}
\centering
\includegraphics[width=8 cm, angle=270]{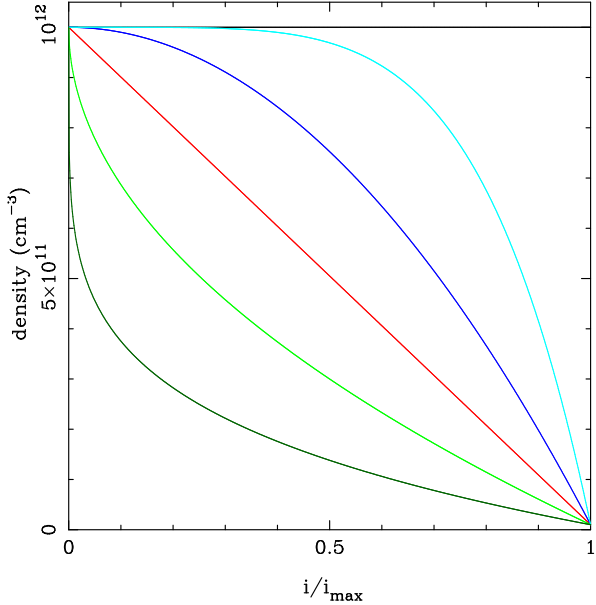}
\caption{{\small Density as a function of step through the \xsc ~chain. The black line corresponds to constant density; the red line corresponds to linearly decreasing density. The green and blue lines are $q$-law profiles, similar in form to those used for the velocities in Figure~\ref{vlaw}. The green lines correspond to $q$=0.2 \& 0.5 (dark and light, respectively) and the blue lines correspond to $q$=2 \& 5 (again, dark and light, respectively). In each case, the initial density is 10$^{12}$ cm$^{-3}$ and, in the non-constant density cases the outer-edge density is 10$^{10}$ cm$^{-3}$.}}
\label{dlaw}
\end{minipage}
\end{figure}

\section{A composite column}
\label{4}

In the winds we have simulated so far, the ionization parameter does not change significantly through the wind material unless the column densities are extremely large. However the gas properties (most notably the density structure - see Section~\ref{3-2}) in these calculations may not be an accurate description of the real physical conditions in an AGN wind. In the more case of a geometrically thick wind, where either the turbulent velocity, or the outflow velocity shear, is greater than the sound speed in the gas, a wide range of ionization states can be present throughout the material.

The variation in the ionization parameter through such a composite column depends on the velocity and density structure of the wind (Section~\ref{2-3}) and on the increasing distance of the gas from the source of the ionizing radiation. We investigate the spectra from a range of composite columns by imposing a range of density profiles on the wind material. Along with a simple constant density profile, we examine density profiles similar in form to the velocity profiles presented earlier. With the exception of the constant density case, the density of shell $i$ in the rest frame of the observer is given by:

\begin{flushright}
\begin{minipage}{80mm}
\begin{eqnarray}
n_{i} & = & n_{l} + (n_{o}-n_{l})\left(\frac{i}{i_{max}}\right)^{q}
\label{eq6}
\end{eqnarray}
\end{minipage}
\end{flushright}

\hspace{-6mm}where $n_{l}$ and $n_{o}$ are the launch and terminal densities, respectively, and $i_{max}$ is the number of steps in the \xsc ~chain. We note that, in a similar vein to the velocity profiles presented earlier, there is no specific physical motivation for any of the density profiles used here. Rather, the resulting spectra highlight the general impact that a composite column has on the absorbed X-ray spectrum of AGN.

Along with the constant density case, we calculate model spectra for five $q$-law density profiles ($q$ = 5.0, 2.0, 1.0, 0.5, 0.2), shown in Figure~\ref{dlaw}. Each profile, with the exception of the constant density case, has a launch density of 10$^{12}$ cm$^{-3}$ and a terminal density of 10$^{10}$ cm$^{-3}$.

\begin{figure*}
\centering
\begin{minipage}{170 mm} 
\centering \includegraphics[width=119 mm, angle=270]{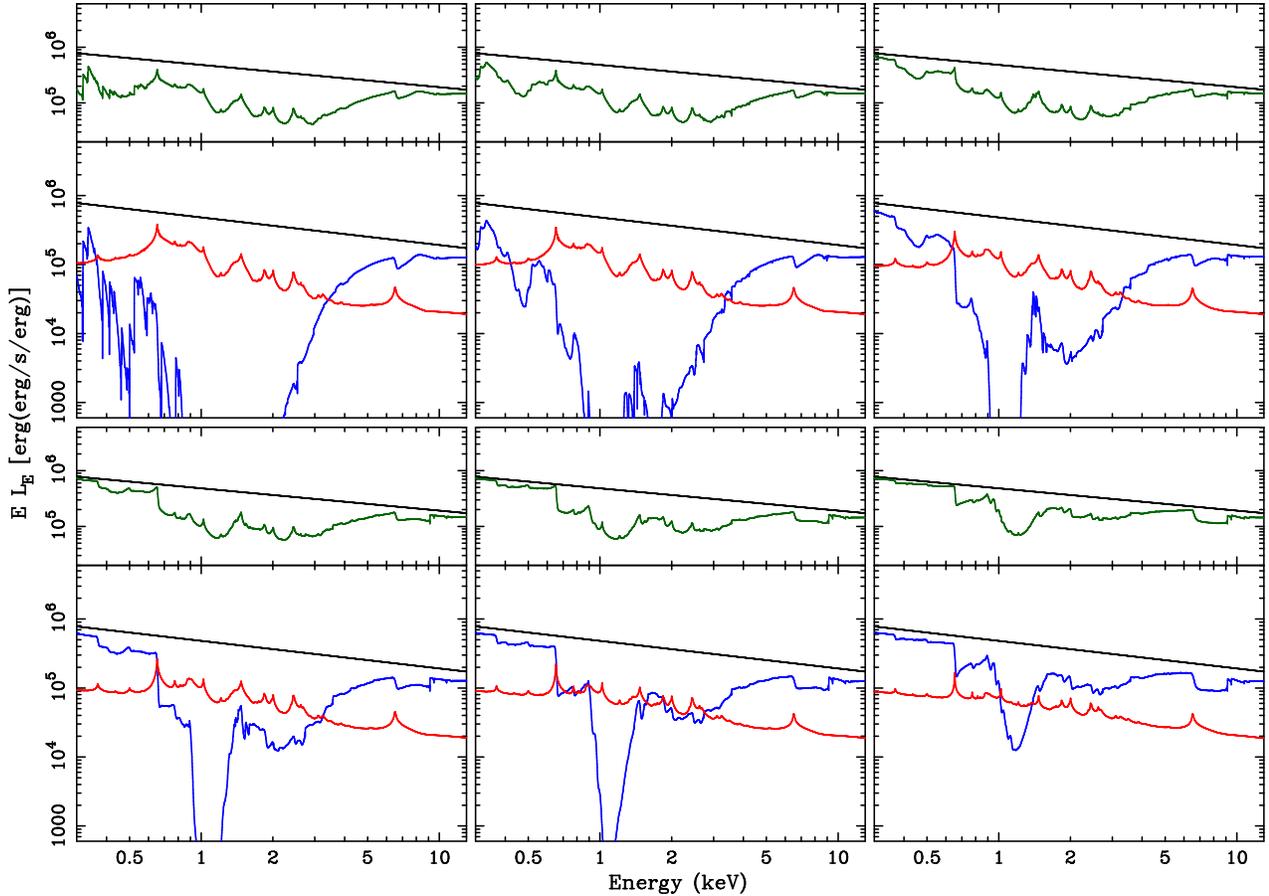}
\caption{{\small As for Figure~\ref{vlawspect} but for \xsc ~runs with different density profiles: Panels 1-3 (Top, left to right): constant density, $q$ = 5.0 \& 2.0. Panels 4-6 (Bottom, left to right): $q$ = 1.0, 0.5 \& 0.2. All other parameters are fixed at the standard values.}}
\label{dspect}
\end{minipage}
\end{figure*}

Figure~\ref{dspect} (panels 1-6) shows the model X-ray spectra from \xsc ~with these density profiles. A comparison of Figure~\ref{dspect}, panel 1, with the X-ray spectrum from the standard set of wind parameters (Figure~\ref{vlawspect}, panel 3), clearly highlights the impact that composite columns have on the X-ray spectrum. In the constant density case, the ionization parameter drops by a factor of $\sim$4 across the wind due to the reduction in luminosity of the source due to relativistic beaming which is now not matched by a change in density to keep the ionization parameter constant. Thus the constant density wind includes material at lower ionization state, giving considerably more absorption on the transmitted spectrum than in the standard case. The change in the ion populations is particularly evident in the shape of the iron K absorption line, where the absence of high ionization species at high velocity results in a more `edge-like' shape compared to the rectangular notch in the standard spectrum.

Panels 2-6 show the changes in spectral shape that result from strongly decreasing density profiles with different $q$-values. In these cases the ionization structure is more complex than in either the standard spectrum case (where the ionization state remains roughly constant throughout the wind), or the constant density case. In the two high-$q$ cases, the density remains high for a considerably amount of the total column. While the density remains high the ionization parameter drops (in a similar fashion to the constant density case), however once the density drops considerably ({\bf O}$\sim$20\%) the ionization state of the material becomes dominated by the density of the material and the ionization parameter begins to rise. The final ionization state at the outer edge of the cloud is higher than the initial ionization state. In the most extreme high-$q$ case shown, the ionization state rises by a factor of $\sim$4.5. In the low-$q$ cases, the density drops rapidly and ionization state of the material is dominated by the dropping density throughout the wind, rising rapidly. In the most extreme low-$q$ case shown the ionization state at the far edge of the wind is $\sim$24 times higher than the initial ionization state.

The spectra from composite columns are, in fact, very similar to the spectra calculated with (approximately) uniform ionization state (compare Figure~\ref{dspect} with Figure~\ref{chispect}), demonstrating that complex ionization structure columns can be modelled with more simple models that have uniform ionization state (at least to first order). The best-fit ionization state from these fits represents an average for the ionized material.

\section{Conclusions}
\label{5}

We adopt a numerical computational method that links a series of radiative transfer calculations, including the effect of a global velocity field in a self-consistent manner (\xsc), in order to produce a considerably more realistic picture of the X-ray spectrum of an AGN seen through an accretion disk wind.

We present a series of example spectra from the \xsc ~code that allow us to examine the shape of AGN X-ray spectra seen through a wind for a range of velocity and density distributions, total column densities and initial ionization parameters. The detailed spectral models presented here clearly show considerable complexity and detailed structure that is strongly affected by all these factors. Nowhere is the impact of these key parameters more clearly highlighted than in the region around iron K$\alpha$. In particular, we use this region to highlight the changes in shape of the absorption and emission features as a function of the assumed velocity law (see Figure~\ref{vironedge}).

The complexity and structure of the \xsc ~models is not present in the ad-hoc spectral models, presented in GD04 \& SD06. The presence of sharp features in the \xsc ~spectra contrasts strongly with the previous models and with the smooth nature of the observed X-ray spectra of AGN with soft X-ray excesses. We note, however, that the models presented here have an energy resolution considerably in excess of that of typical CCD detectors available on current X-ray satellites and that the sharp features in these models will be strongly smeared when convolved with an appropriate instrument response (see Figure~\ref{instrresp}). Despite this, the \xsc ~models clearly show considerable differences in shape in comparison with the previous smeared ionized absorption models. In particular, the \xsc ~spectra have a strong, line-like, feature at $\sim$1.4 keV. This feature is, in fact, not a line (although there is a considerable contribution to the flux at these energies from line emission), it is primarily the result of considerable absorption at energies directly above and below 1.4 keV, associated with ionized oxygen and iron respectively.

Although the current version of the \xsc ~code has some unaddressed issues, it is extremely unlikely that the presence of these strong features will be dramatically affected by addressing these points. The most significant parameter affecting the presence of these features is the terminal velocity that the wind reaches. Increasing the terminal velocity, and hence increasing the velocity dispersion across the wind, will extend the region of the spectrum that is directly impacted upon by the ionized oxygen and iron absorption. Thus, a terminal velocity of $\sim$c could, potentially, result in a spectrum somewhat closer to the ad-hoc smeared, ionized absorption models, presented in GD04 \& SD06, that are known to fit the observed data well. Such fast moving material cannot be associated with a radiatively driven accretion disk wind, however such high velocity material may exist in a radio jet. 

\section{Acknowledgements}

The authors thank Tim Kallman for writing and supporting the \xst ~code, Marek Gierli\'nski, David Wake \& Malgosia Sobolewska for {\small PERL} support, useful discussions and comments.  This research has made extensive use of NASA's Astrophysics Data System Abstract Service.  NJS and CD acknowledge financial support through a PPARC PDRA and Senior fellowship, respectively.

\appendix

\section{\xst ~details}
\label{app1}

The inner workings of \xst ~have been presented elsewhere, notably in the \xst ~manual$^{*}$\footnotetext{\hspace{-2mm}$^{*}$\hspace{1mm}http://heasarc.gsfc.nasa.gov/docs/software/xstar/docs/html/ xstarmanual.html}, however here we reiterate several of the more subtle and important details that effect the \xsc ~code.

\subsection{The propagated spectrum}
\label{app1-1}

The most significant issue concerning \xst s internal processes is the spectrum \xst ~propagates internally. The output transmitted and emitted, spectra from a single \xst ~run (denoted L$^{(5)}_{\epsilon}$, L$^{(6)}_{\epsilon}$ \&  L$^{(7)}_{\epsilon}$, respectively, in the \xst ~manual [chapter 8, section E.4]) are not the same quantity that \xst ~propagates between sub-shells in order to calculate the ionization balance and thermal equilibrium. Using the same notation as the \xst ~manual, the local radiation field propagated from sub-shell $i$ into sub-shell $i+1$ is defined as:

\begin{flushright}
\begin{minipage}{80mm}
\begin{eqnarray}
L^{1}_{\epsilon,i} & = & L^{1}_{\epsilon,i-1}e^{-(\tau_{bf,i}+\tau_{ff,i})} \nonumber\\
& & +\,L_{\epsilon,bf,i}\,+\,L_{\epsilon,ff,i}
\label{aeq1}
\end{eqnarray}
\end{minipage}
\end{flushright}

\hspace{-6mm}where L$^{1}_{\epsilon,i}$ is the specific luminosity output from the sub-shell, $i$ (and hence the input to shell $i+1$) L$^{1}_{\epsilon,i-1}$ is the specific luminosity output from the previous, $i-1$ sub-shell (and thus is the input to shell $i$ - in the case of the initial shell, $i$=1, this is equal to the input incident spectrum, $L_{\epsilon,0}$), $\tau_{bf,i}$ \& $\tau_{ff,i}$ are the optical depths of bound-free and free-free interactions in the sub-shell $i$, respectively, and $L_{\epsilon,bf,i}$ \& $L_{\epsilon,ff,i}$ are, correspondingly, the specific luminosities emitted from sub-shell $i$ due to bound-free and free-free interactions. By contrast, the final output spectra from a completed \xst ~run are given by:

\begin{flushright}
\begin{minipage}{80mm}
\begin{eqnarray}
L^{5}_{\epsilon} & = & L_{\epsilon,0}\,e^{-({\small \Sigma}_{i}(\tau_{bf,i}+\tau_{ff,i}+\tau_{bb,i})\,)}\label{aeq2}\\
L^{6}_{\epsilon} & = & \sum_{i}\,(L_{\epsilon,bb,i}\,+\,L_{\epsilon,bf,i}\,+\,L_{\epsilon,ff,i})\,\mid\,_{inward}\label{aeq3}\\
L^{7}_{\epsilon} & = & \sum_{i}\,(L_{\epsilon,bb,i}\,+\,L_{\epsilon,bf,i}\,+\,L_{\epsilon,ff,i})\,\mid\,_{outward}\label{aeq4}
\end{eqnarray}
\end{minipage}
\end{flushright}

\hspace{-6mm}where $\tau_{bb,i}$ if the optical depth of bound-bound transitions in sub-shell $i$. Since we require consistency with a single \xst ~run, we must propagate the same thing that \xst ~propagates internally, between our linked \xst ~runs. Unfortunately, it is not simple to reconstruct the locally propagated radiation field from the final sub-shell, $i_{max}$, of an \xst ~run (L$^{1}_{\epsilon,i_{max}}$) using the output products of that \xst ~run (equations~\ref{eq2},~\ref{eq3} \&~\ref{eq4}). Instead, we modify the \xst ~source code directly to output L$^{1}_{\epsilon,i_{max}}$ for a given \xst ~run, allowing us to use this quantity as the input spectrum for the next \xst ~run in the chain.

\subsection{Ionization parameter}
\label{app1-2}

Internally, \xst ~calculates the ionization state for each sub-shell based on the propagated specific luminosity from the previous sub-shell, the density of the sub-shell and the thickness of the sub-shell. Although possible in principle, we do not specify the radial profile of the ionization parameter for the \xsc ~spectra presented here. Instead, we follow a similar process to the internal \xst ~calculation. The initial ionization parameter is specified at the innermost edge of the wind material and then the ionization parameter input to each link in the chain is calculated as a function of distance and absorption along the l.o.s for each shell, $i$:

\begin{flushright}
\begin{minipage}{80mm}
\begin{eqnarray}
\xi_{i} & = & \frac{L_{i}}{n_{i} R_{i}^{2}}
\label{aeq5}
\end{eqnarray}
\end{minipage}
\end{flushright}

\hspace{-6mm}where L$_{i}$ is the luminosity propagated into shell $i$ ({\it i.e.} the luminosity of the specific luminosity spectrum propagated into shell $i$, L$^{1}_{\epsilon,i-1}$, calculated over the 1-1000 Rydberg range - see Equation~\ref{aeq1}), $n_{i}$ is the rest frame gas density in shell $i$ and $R_{i}$ is the radius of shell $i$ from the central X-ray source. $R_{i}$ is given by:

\begin{flushright}
\begin{minipage}{80mm}
\begin{eqnarray}
R_{i} & = & R_{l}+\sum_{j=1}^{i-1}\left(\frac{N_{H,j}}{n_{j}}\right) 
\label{aeq6}
\end{eqnarray}
\end{minipage}
\end{flushright}

\hspace{-6mm}where $R_{l}$ is the innermost launch radius of the wind, N$_{H,j}$ is the column density of shell $j$ and $n_{j}$ is the density of the wind material in shell $j$.

The radial dependence of the ionization parameter is largely unimportant here since the thickness of each shell in the winds simulated is much less than the launch radius of the wind, as is the thickness of the wind as a whole. The radial dependence of the ionization parameter will, however, be important for physical scenarios where the wind may have considerable radial thickness.

\subsection{Velocity Transforms}
\label{app1-3}

Velocity transforms are used calculate both the spectrum propagated between each shell in the chain, and the transmitted and emitted spectrum in the observers frame. In the case of a redshift the transformed specific luminosity spectrum is given by:

\begin{flushright}
\begin{minipage}{80mm}
\begin{eqnarray}
L_{\epsilon}^\prime & = &  L_{\epsilon}\gamma^{3}(1+v/c)^{3}
\label{aeq7}
\end{eqnarray}
\end{minipage}
\end{flushright}

\hspace{-6mm}where $L_{\epsilon}$ is the original specific luminosity spectrum, $\gamma=(1-v^{2}/c^{2})^{1/2}$ and $v$ is the relative radial velocity of the two reference frames.

In the case of a blue-shift, the transformed specific luminosity spectrum is given by:

\begin{flushright}
\begin{minipage}{80mm}
\begin{eqnarray}
L_{\epsilon}^\prime & = &  L_{\epsilon}\left(\frac{1}{\gamma(1-v/c)}\right)^{3}
\label{aeq8}
\end{eqnarray}
\end{minipage}
\end{flushright}
 
\subsection{Velocity shear characterization}
\label{app1-4}

We assume that the velocity shear across each individual shell is linear and we use the standard deviation of this distribution as a representation of the actual velocty shear. The mean velocity in shell $i$ is given by:

\begin{flushright}
\begin{minipage}{80mm}
\begin{eqnarray}
<v_{rad,i}> & = & \frac{1}{2}(v_{rad,i-1}+v_{rad,i})
\label{aeq9}
\end{eqnarray}
\end{minipage}
\end{flushright}

The mean squared velocity is given by integrating the velocity distribution as a function of column density through the shell (we note that for the following equations we have removed the subscript `$rad$' on the velocity terms for clarity purposes):

\begin{flushright}
\begin{minipage}{80mm}
\begin{eqnarray}
\hspace{-4mm}<v_{i}^{2}> & \hspace{-3mm}= & \hspace{-3mm}\frac{\int^{N_{H,i}}_{0}\,\left(v_{i-1}+(v_{i}-v_{i-1})\left(\frac{N_{H}}{N_{H,i}}\right)\right)^{2}dN_{H}}{\int^{N_{H,i}}_{0}\,dN_{H}}\label{aeq10}\\
\hspace{-4mm} & \hspace{-3mm}= & \hspace{-3mm}\frac{1}{3}((v_{i}-v_{i-1})^{2}+3v_{i}v_{i-1})\label{aeq11}
\end{eqnarray}
\end{minipage}
\end{flushright}

\hspace{-6mm}where $N_{H,i}$ is the total column density of the shell $i$. The standard deviation is given by the standard equation $\sigma_{i}^{2}$=$<v_{rad,i}^{2}>-<v_{rad,i}>^{2}$ and substituting Equations~\ref{aeq9} \& ~\ref{aeq11} into this expression thus gives:

\begin{flushright}
\begin{minipage}{80mm}
\begin{eqnarray}
\sigma_{i} & = & (v_{rad,i}-v_{rad,i-1})/\sqrt{12}
\label{aeq12}
\end{eqnarray}
\end{minipage}
\end{flushright}

 The turbulent velocity in the \xst ~run for shell $i$, v$_{turb,i}$ is then set to be equal to this standard deviation, but with an additional component, $v_{t}$, representing the intrinsic, real, turbulent velocity of the material (see Equation~\ref{eq1}).

\section{Numerical Limitations}
\label{app2}
 
Here we note some of the technical limitations of the \xsc ~simulation process. The number of steps in the chain and interpolation routines used for each of the velocity shifts, along with, the number of area segments in the emission spectrum calculation, all present numerical limitations in the accuracy of the output spectrum.

The most significant of these concern the number of steps in the chain and the interpolation errors associated with the velocity shifts. When the number of steps in the chain is small ($<$40 runs), the velocity grid is coarse and sinusoidal wiggles are introduced to the spectrum, particularly around discrete features. Conversely, when the number of steps in the chain is high ($>$150 runs), the cumulative errors from the large number of interpolations run over the course of the chain (three for each step) introduce significant changes in spectral shape of the output spectrum. In particular, when the width of narrow features in the spectrum is comparable to the size of the energy grid, a large number of interpolations fills in the absorption features, and reduces the strength of emission features. In extreme cases, these errors can be very significant (factors of $\sim$3-4) but they can be minimised either by ensuring that the turbulent velocities used throughout the chain are always large enough that none of the spectral features are narrow in comparison with the energy grid, or by increasing the density of the energy grid prior to interpolation. The latter solution will be implemented in future versions of the code.

By contrast, the number of area segments used in the emission spectrum calculation introduces relatively minor errors to the calculated spectrum. Using a small number of area segments ($<$200) results in discrete wiggles in the emission spectrum, particularly on the blue-ward side of strong emission features. Using a large number of area segments ($>$5000) rapidly becomes prohibitively computationally intensive.

\end{document}